\documentclass[dvips]{acta}
\usepackage{supertabular,lscape,epsfig}
\usepackage{amssymb}
\usepackage{amsmath}
\DeclareSymbolFont{ppa}{OT1}{ppl}{m}{it}
\DeclareMathSymbol{\vv}{\mathalpha}{ppa}{'166}

\newfont{\hb}{rphvb at 10pt}
\newfont{\hbo}{rphvbo at 10pt}
\newfont{\bitt}{rptmbi at 12pt}
\newfont{\bits}{rptmbi at 11pt}

\SetPages{53}{66}

\SetVol{63}{2013}

\begin{document}

\newcommand{\TabApp}[2]{\begin{center}\parbox[t]{#1}{\centerline{
  {\bf Appendix}}
  \vskip2mm
  \centerline{\small {\spaceskip 2pt plus 1pt minus 1pt T a b l e}
  \refstepcounter{table}\thetable}
  \vskip2mm
  \centerline{\footnotesize #2}}
  \vskip3mm
\end{center}}

\newcommand{\TabCapp}[2]{\begin{center}\parbox[t]{#1}{\centerline{
  \small {\spaceskip 2pt plus 1pt minus 1pt T a b l e}
  \refstepcounter{table}\thetable}
  \vskip2mm
  \centerline{\footnotesize #2}}
  \vskip3mm
\end{center}}

\newcommand{\TTabCap}[3]{\begin{center}\parbox[t]{#1}{\centerline{
  \small {\spaceskip 2pt plus 1pt minus 1pt T a b l e}
  \refstepcounter{table}\thetable}
  \vskip2mm
  \centerline{\footnotesize #2}
  \centerline{\footnotesize #3}}
  \vskip1mm
\end{center}}

\newcommand{\MakeTableApp}[4]{\begin{table}[p]\TabApp{#2}{#3}
  \begin{center} \TableFont \begin{tabular}{#1} #4 
  \end{tabular}\end{center}\end{table}}

\newcommand{\MakeTableSepp}[4]{\begin{table}[p]\TabCapp{#2}{#3}
  \begin{center} \TableFont \begin{tabular}{#1} #4 
  \end{tabular}\end{center}\end{table}}

\newcommand{\MakeTableee}[4]{\begin{table}[htb]\TabCapp{#2}{#3}
  \begin{center} \TableFont \begin{tabular}{#1} #4
  \end{tabular}\end{center}\end{table}}

\newcommand{\MakeTablee}[5]{\begin{table}[htb]\TTabCap{#2}{#3}{#4}
  \begin{center} \TableFont \begin{tabular}{#1} #5 
  \end{tabular}\end{center}\end{table}}

\newfont{\bb}{ptmbi8t at 12pt}
\newfont{\bbb}{cmbxti10}
\newfont{\bbbb}{cmbxti10 at 9pt}
\newcommand{\uprule}{\rule{0pt}{2.5ex}}
\newcommand{\douprule}{\rule[-2ex]{0pt}{4.5ex}}
\newcommand{\dorule}{\rule[-2ex]{0pt}{2ex}}
\begin{Titlepage}
\Title{ASAS Photometry of ROSAT Sources.\\
II.~New Variables from the ASAS North Survey}
\vspace*{-3pt}
\Author{M.~~K~i~r~a~g~a~~~and~~~K.~~S~t~ê~p~i~e~ñ}{Warsaw University
Observatory, Al.~Ujazdowskie~4, 00-478~Warsaw, Poland\\
e-mail: (kiraga,kst)@astrouw.edu.pl}
\vspace*{-3pt}
\Received{January 22, 2013}
\end{Titlepage}

\vspace*{-3pt}
\Abstract{We present a catalog of 307 optical counterparts of the bright
ROSAT X-ray sources, identified with the ASAS North survey data and
showing periodic brightness variations. They all have declination north
of $-25\arcd$. Other data available from the literature for the listed stars
are also included. All the tabulated stars are new variables, except for 13
previously known, for which the revised values of periods are given.}{Stars:
variables: general -- Stars: rotation -- Stars: activity -- X-rays: stars}

\vspace*{-3pt}
\Section{Introduction}
Coronal-chromospheric activity is common among cool stars possessing
subphotospheric convection zones (Wilson 1963, Pallavicini \etal 1981). Its
ubiquitous presence in these stars finds a support in theoretical
considerations, according to which the presence of a convective layer in a
rotating body is sufficient for development of the dynamo mechanism
efficiently strengthening even a very weak, ``seed'' magnetic field (Parker
1955). The field is most likely generated at the interface between the
convection zone and the radiative core or, possibly, in the bulk of the
convective layer. Due to buoyancy, magnetic tubes are brought to stellar
surface where they interact with convection, producing several phenomena
known under the collective name of activity. Observations and theory show
that the stellar activity level depends on the internal structure of a star
and its rotation rate (Durney and Latour 1978, Noyes \etal 1984). In
particular, rapidly rotating cool stars show high levels of activity as
measured by strong surface magnetic fields, intense chromospheric and
coronal emission, and appearance of the numerous cool star spots believed
to be regions of the high concentration of the local magnetic field. As the
observations of cool stars (including our Sun) show, the spots are often
distributed nonuniformly over the stellar surface, hence rotational
modulation of the star's brightness is expected. Systematic photometry of a
large number of active stars makes possible the detection of this
modulation and determination of the stellar rotation period -- a crucial
parameter for analyzing the activity level, and its dependence on global
stellar parameters such as mass, age, evolutionary stage or chemical
composition. Ideal for this purpose are photometric surveys covering a
large part of the sky. One of these is ASAS (Pojmañski 1997, 2004). It
began twelve years ago on the Southern hemisphere (ASAS-S) while its
Northern twin (ASAS-N) is in operation since 2006.

The analysis of the rotation of main sequence M-type stars, based on the
ASAS-S data, was carried out by Kiraga and Stêpieñ (2007). Many new
rotation periods were determined, among them a number with the length of
several tens of days, which were particularly useful for discussion of the
period-activity relation. Later, Kiraga (2012) analyzed all stars from the
ASAS-S survey for which the X-ray flux could be found in the ROSAT Bright
Source Catalog (RBSC) from the ROSAT All Sky Survey. The ASAS-S photometry
is available for stars with declination up to $+29\arcd$. Kiraga (2012)
identified 6028 such stars. Among them 2302 stars showed periodic
variability. An extensive catalog was prepared where all the periodic
variables are listed, together with the data on their variability type and
other interesting information available from the literature.

The present paper is a continuation of the investigation by Kiraga (2012).
We repeat the same procedure for the Northern ROSAT bright sources and
photometric observations from the ASAS-N survey as Kiraga (2012) did for
the Southern hemisphere.

\section{X-Ray and Photometric Observations and Data Analysis}
The ROSAT All Sky Survey was obtained between July 30, 1990 and January 25,
1991, with additional observations taken in February and August 1991 (Voges
\etal 1999). Analysis of the survey data resulted in the detection of 145\,060
sources. Among them 18\,811 sources were classified as ``bright'' and
published by Voges \etal (1999) in RBSC. A source was classified as bright
when its count rate was higher than 0.05~cts/s in the 0.1--2.4~keV energy
band and at least 15 photon counts were registered.

The ASAS-N station is located at Haleakala (Maui, Hawaii Islands). The
photometric survey is carried out using the Nikkor 200mm f/2.0 lens
equipped with the Apogee AP-10, $2048\times2048$ CCD camera which has the
angular resolution of 15~arcsec/pixel. We used only the {\it V}-band data
in the present analysis, taken between May 29, 2006 and February 12, 2012
for stars with declination north of $-25\arcd$. We decided to include into
our analysis also stars from the equatorial belt (with declination between
$-25\arcd$ and $+29\arcd$) which had already been investigated by Kiraga
(2012). We did so because ASAS-N has a larger aperture, hence useful data
are available for stars fainter than from ASAS-S. Furthermore, ASAS-S
stopped collecting data in October 2008 due to some problems with the new
CCD cameras so the data obtained later by ASAS-N may be used for
variability search of previously non detected variables.

There are 12\,910 entries in the RBSC located north of declination
$-25\arcd$. The optical identification of the X-ray sources was based on
the coordinates of the X-ray sources given in the ROSAT catalog. Using the
ASAS photometric database we searched for an optical counterpart within
30\arcs from each RBSC source. The distance of 30\arcs corresponds
approximately to two ASAS pixels. We restricted our attention only to stars
with the mean {\it V} brightness between 8~mag and 14.0~mag for which at
least 40 observations are available, so a reasonable period analysis can be
performed.

We identified 4324 stars fulfilling the above criteria. A preliminary
period search was done with the {\sc AoV} algorithm developed and described
by Schwarzen\-berg-Czerny (1989). Because many stars show season to season
variations, usually interpreted as resulting from the long term activity
cycles, we carried out the period search separately for each season, unless
the number of useful observations in the particular season was lower than
40. A period search was also done on the whole data set after shifting the
individual seasonal means to the common level. The data points deviating by
more than 3.5 standard deviations from the seasonal mean were treated as
outliers and rejected (Kiraga and Stêpieñ 2007).

Using 6 phase bins, we obtained the value of the {\sc AoV} statistics
higher than 10 for 1603 stars, of which 647 have already been listed in
Kiraga (2012). The latter stars were excluded from the further analysis.
Although the majority of the presently obtained periods are in agreement
with the old ones, several new periods differ substantially, \eg by a
factor of two, or are related to one another as 1~d aliases. There are,
however, stars for which new periods are not apparently related to the old
ones. Detailed comparison of the new and old periods as well as a
discussion of all detected variables in the equatorial belt covered by both
ASAS surveys will be presented in the forthcoming paper.
 
Of the remaining 956 objects, 37 were found in the ASAS catalog ACVS and
additional 159 stars were listed in the catalogs by Norton \etal (2007),
Hartman \etal (2011) and Pigulski \etal (2009). So, after the first step,
we were left with 760 suspected periodic variables. In the second step, we
applied the CLEAN algorithm (Roberts \etal 1987) to each data set and, in
addition, we critically inspected the resulting photometric curves
visually. Many of the stars which had passed the CLEAN test, are marked as
variables in the SIMBAD database. They were also removed from our sample,
unless a new or revised value of the period was obtained with a high level
of confidence (see below). As a final step, we formed a list of 307 objects
with definitely present periodic variability.

Although the approximate estimate indicates that out of 4000 coincidences
between X-ray and ASAS sources about 0.5\% may be spurious, the probability
of a spurious coincidence of a variable star and the X-ray source is much
lower so we expect that at most one such case occurs among 307 investigated
objects.

So far, only {\it V}-band ASAS-N photometric data are reduced and released
to a public domain. In lack of measurements in another optical band, we
used 2MASS photometry (Skrutskie \etal 2006) to obtain color information
for each investigated star. We adopted the {\it J}-band data because a
reliable calibration of the bolometric correction based on the $V-J$ color
exists for F--M stars (Casagrande \etal 2008, 2010). Since 2MASS survey has
a substantially better resolution than ASAS (2\zdot\arcs5 \vs 15\arcs) we
had to apply a special procedure to obtain a reliable {\it J} magnitude
related to our optical object. Low resolution of the ASAS survey means that
fluxes from all sources lying closer than about 30\arcs from the optically
identified object add to its {\it V} measurement. To apply a respective
correction to the {\it J} data, we counted all the 2MASS sources within
40\arcs (with allowance for the additional margin of 10\arcs) from the
position of the ASAS source and added together their {\it J} fluxes. The
brightest source was assumed to correspond to the optical object but its
{\it J} brightness was corrected by adding to it the fluxes of all the
counted nearby sources. The directly measured {\it J} magnitude of the
brightest source is listed in our catalog, together with the relative
total {\it J} flux of all the counted sources, expressed in units of the
brightest source flux. A relative total flux like 1.04 means that the
adopted {\it J} flux of the brightest source was corrected by 4\% before
forming a color index $V-J$. As it is seen from the catalog, the relative
total flux does not exceed 1.05 in most cases but it can reach a value of
2 or more for a star with nearby companion(s) of a comparable
brightness. The total flux expressed in magnitudes is listed as $J_{\rm
cor}$. This flux was used together with the {\it V}-magnitude to
calculate the bolometric correction of each identified star. For stars
with $(V-J_{\rm cor})<2.4$~mag the formula given by Casagrande \etal (2010)
was used (where $F_{\rm bol}$ is expressed in erg cm$^{-2}$\,s$^-1$):
$$\log(F_{\rm bol})=-0.4\cdot J+\log(fc)-5$$ 
where 
$$fc=2.7915-2.8096\cdot(V-J_{\rm cor})+1.2799\cdot (V-J_{\rm cor})^2-0.2049\cdot (V-J_{\rm cor})^3.$$

For stars with $(V-J_{\rm cor})>2.4$~mag we used the second order formula
obtained from the fitting to the data given by Casagrande \etal (2008):
$$BC_J=0.8439+0.3652\cdot(V-J_{\rm cor})-0.0267\cdot(V-J_{\rm cor})^2,$$
$$m_{\rm bol}=J+BC_{J_{\rm cor}},$$ 
and 
$$\log(F_{\rm bol})=-0.4\cdot(m_{\rm bol}-4.75)-6.489.$$

The uncertainty of the bolometric flux is dominated by the uncertainty of
the {\it J} and $J_{\rm cor}$ because it is based on a single measurement
taken at a random phase of the variability period. We assume that it is
equal to the standard deviation of the {\it V} measurement for each star
\ie $\sigma_V$. Other uncertainties, like the error of $\langle V\rangle$,
or calibration errors are significantly lower. According to our estimate,
the error of $\log(F_{\rm bol})$ does not exceed 0.05 (but see below).

The X-ray flux in the 0.1--2.4~keV range was calculated from the data given
in RBSC using a formula provided by Fleming \etal (1995):
$$F_x=(5.31{\it HR1}+8.31)\cdot10^{-12}~{\rm cts}$$
where {\it HR1} is the hardness ratio:
$${\it HR1}=(B-A)/(A+B)$$
with $B$ -- a number of counts in the 0.4--2.0~keV range and $A$ -- a
number of counts in the 0.1--0.4~keV range. Here $F_x$ is expressed in
erg cm$^{-2}$\,s$^{-1}$ and ``cts'' means counts per second.

From $F_x$ and $F_{\rm bol}$ the ratio $R_x=\log(F_x/F_{\rm bol})$ was
calculated for each variable. The error of $R_x$ is dominated by the error
of the X-ray measurement. It is listed in the catalog along with the value
of $R_x$. However, for stars with large relative total fluxes, the $J_{\rm
cor}$-value, hence $\log(F_{\rm bol})$, may be erroneous, due to a
possible misidentification of the optical source or the presence of the
nearby, bright infrared source. So, the values of $R_x$ for stars with the
relative total flux exceeding, say, 1.25 should be treated with caution.

\section{New Variable Stars}
\subsection{Properties of New Variables}
In this section we briefly summarize the basic properties of the new
variable stars. We also call attention to several individual stars with
particularly interesting properties.

Most of the investigated stars have so far not been extensively observed so
very limited data on them can be found in the SIMBAD database. Spectral
types are known only for 136 stars, parallaxes have been measured for 17,
and radial velocity for mere 19 objects. Among those with known spectral
classification there are two A-type, 12 F-type, 35 G-type, 73 K-type and 14
M-type stars.

The classification of the variability type of the new variables was based
on the visually identified characteristic features of their light curves,
and on the level of X-ray emission relative to the bolometric flux. As a
result, 266 stars were classified as spotted rotators, although a low
amplitude variation in some of them makes difficult to distinguish a
spotted variable from an ellipsoidal, or tidally deformed object. Among
other variables 10 contact binaries were identified, 12 eclipsing binaries
with deformed components but not in contact and 18 detached eclipsing
binaries. One optical counterpart (ASAS 071404+7004.3) to the X-ray source
1RXS J071404.0+700413 was classified as a miscellaneous variable.

\begin{figure}[htb]
\centerline{\includegraphics[width=13.5cm]{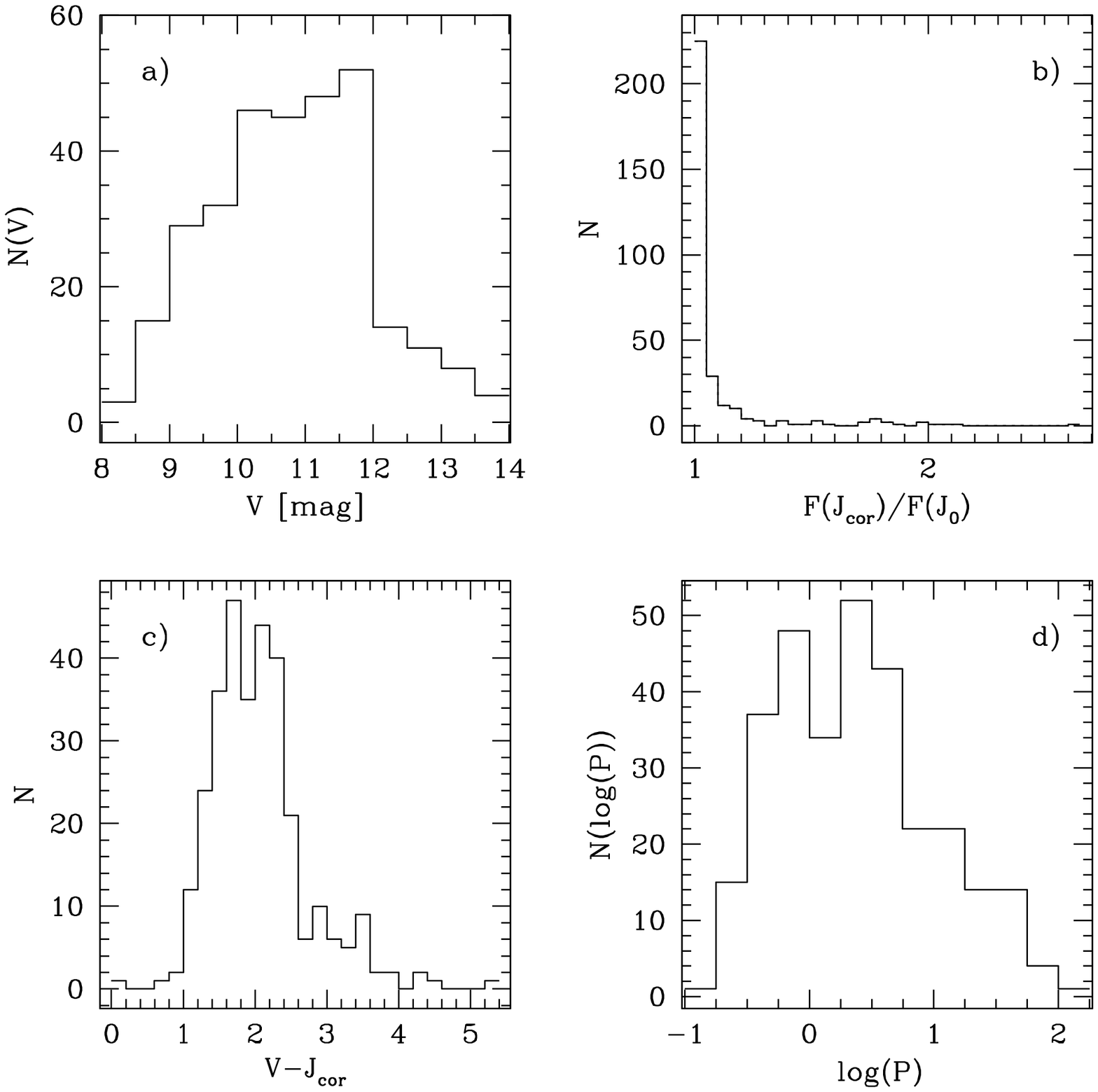}}
\FigCap{{\it a)}~Apparent {\it V}-distribution of the new variables in
0.5~mag bins.
{\it b)}~Histogram of the flux in the {\it J}-band from all sources located
within 40\arcs of ASAS position expressed in the {\it J}-band flux from
the brightest source (see text).  For most stars the contribution to {\it
J}-band flux from fainter objects is small.
{\it c)}~Color index $V-J_{\rm cor}$ distribution of the new variables in
0.1~mag bin. See text for definition of $J_{\rm cor}$.
{\it d)}~Period distribution of the detected variables. The number of stars
is given per 0.25 bin in $\log(P)$ (with period in days).}
\end{figure}

Fig.~1 presents a few statistical properties of the investigated stars.
Fig.~1a gives an apparent magnitude distribution of the new variables. The
number of stars increases almost linearly between 9~mag and 12~mag and falls
off outside of this range. In our sample, there are only three stars brighter
than 8.5~mag, of which ASAS 025746+4431.5 (HD 18281) with $V=8.32$~mag is
the brightest, and five stars fainter than 13.5~mag. The histogram
representing the number of stars as a function of the relative total flux
is presented in Fig.~1b. This flux is lower than 1.05 for 225 stars, and
exceeds 1.25 for 27 stars.  Among the latter, it is higher than 2.0 for four
stars. Fig.~1c shows the color distribution of the variables. Most of them
fall within the range $1<(V-J_{\rm cor})<3$. Only four stars are hotter than
that and 28 stars have the color index exceeding 3.0. The period
distribution is shown in Fig.~1d. About 90\% of stars are in the period
range between 0.315~d and 37~d with overall maximum of the period
distribution around 2--3~d. Among the short-period variables two have
periods shorter than 0.2~d and three have periods between 0.2~d and
0.25~d. In the long-period tail two stars have periods longer than 100~d
and four have periods between 50~d and 100~d.

\subsection{Comments on a Few Interesting Stars}
\hglue-3pt ASAS 071404+7004.3 is the hottest star in our sample with $(V{-}J_{\rm cor}){=}
0.09$~mag which seems to rule out the coronal origin of its X-ray
flux. This star has also an unusually high X-ray to bolometric flux ratio
$R_x= -2.46$, and the light curve very different from other analyzed stars
(see Fig.~2). It may be a previously unrecognized cataclysmic variable with
a period of 22.4~d.
\begin{figure}[htb]
\centerline{\includegraphics[width=13cm, bb=44 565 565 705]{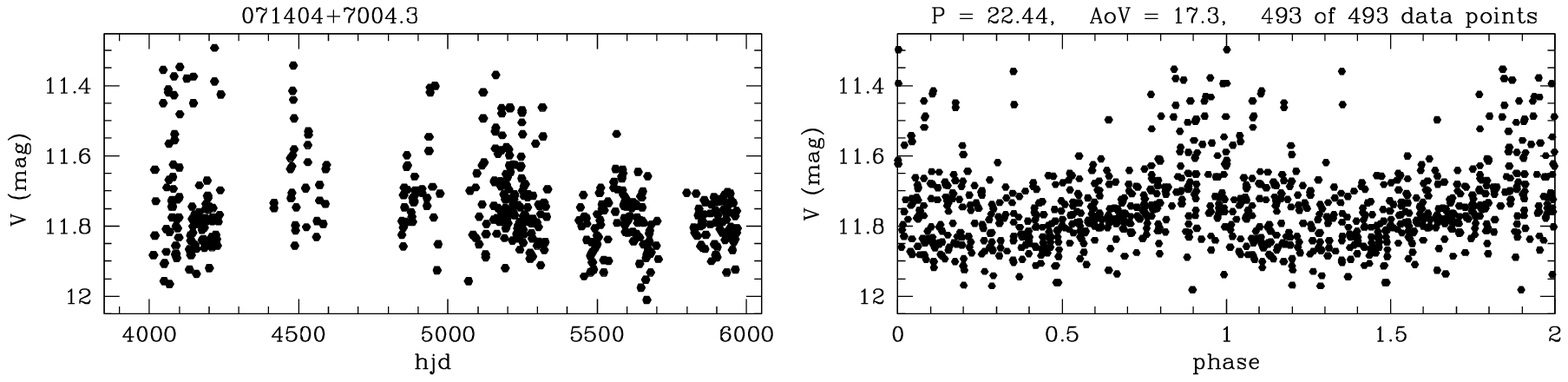}}
\FigCap{Photometric data points for ASAS 071404+7004.3 are presented on the
{\it left panel} (as a function of ${\rm hjd=HJD-2450000}$). Phased light
curve of data showing periodic variability is presented on the {\it right
panel}.}
\end{figure}

ASAS 205846+1417.8 (HD 199742) has the reddest color in our sample,
($V-J_{\rm cor})=5.35$~mag, and has also the highest bolometric
luminosity. Its small proper motion suggests that it may be a distant
M-type red giant, however, its variability period of only 22~d seems to be
too short for the rotational modulation. The star may also be an example of
a low amplitude pulsating red giant (Soszyñski \etal 2011) although these
stars are not known to possess active coronae.

ASAS 235257+6250.0 (BD+62 2316) has the shortest period of
$P{=}0.158038$~d. It is classified as a spotted rotator due to its low
amplitude and variable light curve. Similarly classified is a star (ASAS
034327+2935.6 = BD+29 599, $P=0.187213$) with the second shortest
period. However, the classification of both stars is very uncertain; they
can very well be eclipsing contact (EC) binaries with periods two times
longer than adopted here. Two other stars with period shorter than 0.25~d
are certain EC binaries (ASAS 040959+5558.9 = BD+55 849 -- with
$P=0.242750$~d and ASAS 082931+7231.7 with $P=0.231400$~d).

ASAS 030755+6031.4, with $P=0.20629$~d, shows a very peculiar light curve,
reminiscent that of a pulsating star with a sharp rise and a gradual
decline (see Fig.~3) but we classified it as a rotationally variable due to
its X-ray emission typical of the high activity stars.
\begin{figure}[htb]
\centerline{\includegraphics[width=13cm, bb=45 565 565 705]{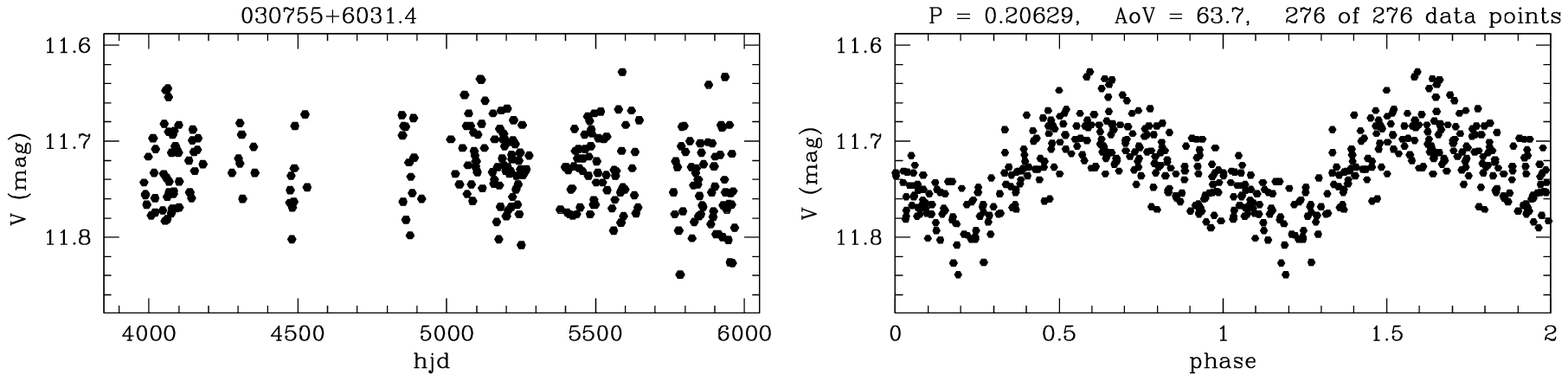}}
\vskip3pt
\FigCap{Photometric data points for ASAS 030755+6031.4 are presented on the
{\it left panel} (as a function of ${\rm hjd=HJD-2450000}$). Phased light
curve of data showing periodic variability is presented on the {\it right
panel}.}
\end{figure}

Periods of 10 EC binaries are between $P{=}0.231400$~d (ASAS 082931+7231.7)
and $P=0.376532$~d (ASAS 034501+4937.0). Detached eclipsing (ED) binaries
have periods from $P{=}0.659984$~d (ASAS 141331+2644.8) to $P{=}3.8581$~d
(ASAS 083103+6948.9), whereas the periods of eclipsing $\beta$-Lyr type
(EB) binaries fall between 0.393273~d (ASAS 061335+4914.1) and 37.685~d
(ASAS 224336+4445.1). The latter star is very likely an ED binary with
narrow eclipses and strong brightness variations between eclipses due to
the presence of spots (see Fig.~4).
\begin{figure}[htb]
\centerline{\includegraphics[width=13cm, bb=45 565 565 705]{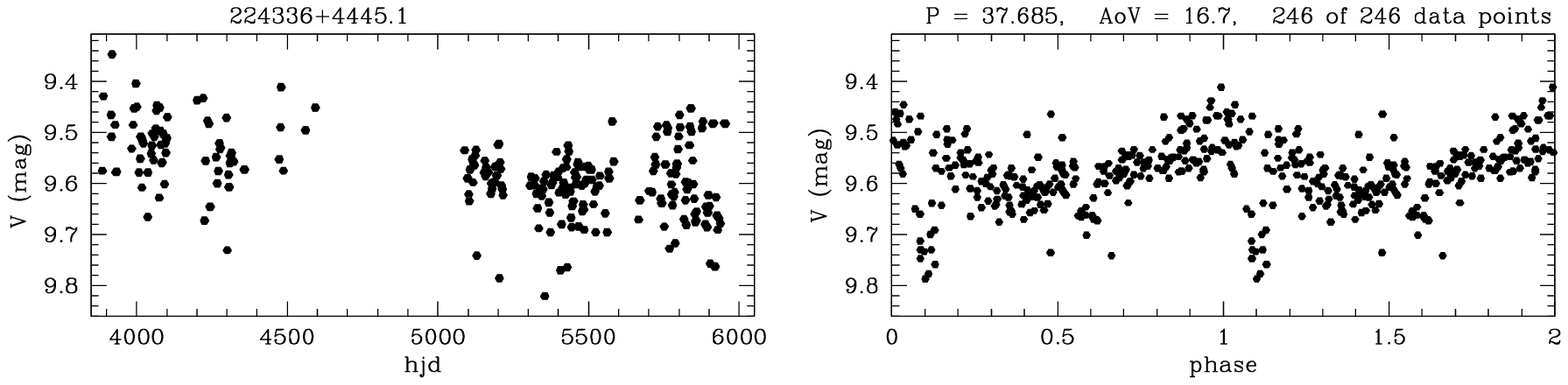}}
\vskip3pt
\FigCap{Photometric data points for ASAS 224336+4445.1 are presented on the
{\it left panel} (as a function of ${\rm hjd=HJD-2450000}$). Phased light
curve of data showing periodic variability is presented on the {\it right
panel}.}
\end{figure}

All stars with periods above 50 d (ASAS 025412+4035.4 -- HD 17930,
$P=100$~d; ASAS 062445+4316.1 -- BD+43 1525, $P=64.8$~d; ASAS 202740+7734.3
-- BD+77 779, $P=56.8$~d; ASAS 203417+3100.6 -- HD 334646, $P=68.9$~d; ASAS
205131+4547.2 -- BD+45 3306, $P=53.8$~d; ASAS 213240+3604.7 -- HD 205173,
$P=109.7$~d) have spectral types in the range G5--K2 and proper motions
below 31~mas/year. Their $R_x$ values are in the range from $-4.3$ to
$-3.6$, and the hardness ratio is above 0.40. All these data suggest that
they are active giants.

\subsection{Data on the Investigated Variables}
\begin{figure}[p]
\vglue-7mm
\hglue-6mm{\includegraphics[width=14.5cm]{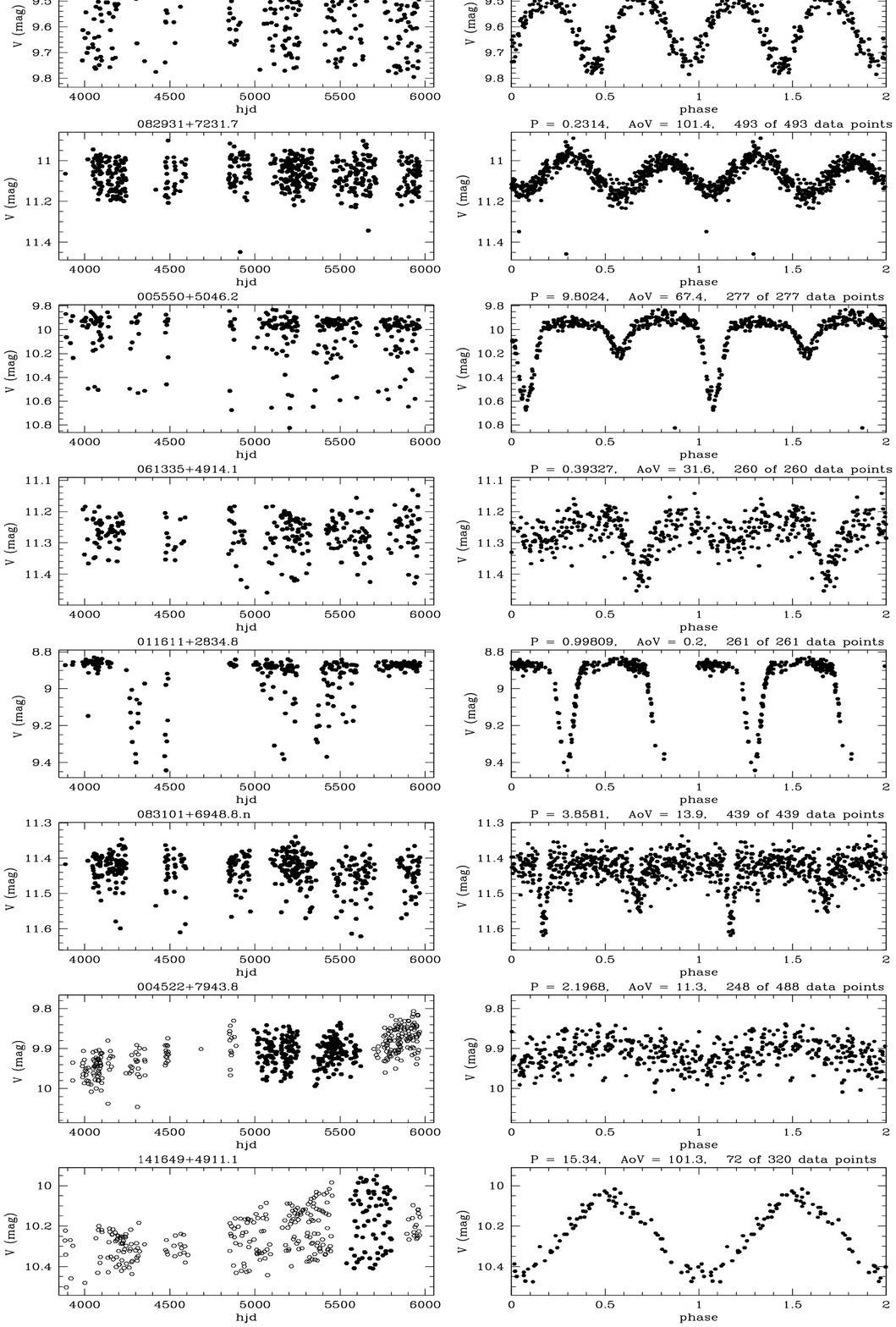}}
\vskip-6pt
\FigCap{A few exemplary light curves of the new variables. {\it Left
panels}: {\it V}-magnitude \vs ${\rm hjd=HJD}-2450000$, {\it right panels}:
phased light curves. From {\it top} to {\it bottom}: two contact binaries,
two $\beta$-Lyr type binary stars, two detached binary stars and two
spotted rotators.}
\end{figure}
\renewcommand{\TableFont}{\scriptsize}
\renewcommand{\arraystretch}{1.1}
\MakeTable{c@{\hspace{7pt}}c@{\hspace{3pt}}c@{\hspace{3pt}}
c@{\hspace{3pt}}c@{\hspace{1pt}}c@{\hspace{3pt}}c@{\hspace{3pt}}c}{12.5cm}{First three entries of Table~1}
{\hline
col. 1          & col. 2             & col. 3      & col. 4         & col. 5             & col. 6                     & col. 7             & col. 8\\\hline
1RXS            & ASAS               & dist        & other          & name               & sp. type                   & $\mu_\alpha$ [mas] &$\mu_\delta$ [mas]\\\hline
002116.4+505315 & 002116+5053.4      & 11.6        & TYC            & 3259-409-1         & $\sim$                     & 29.3               &$-9.2$\\
002415.6+603454 & 002415+6035.0      &  7.5        & 1RXS           & J002415.6+603454   & $\sim$                     &  0.0               & 0.0\\
003327.5+575045 & 003329+5750.6      & 15.8        & 1RXS           & J003327.5+575045   & $\sim$                     &  0.0               & 0.0\\
                &                    &             &                &                    &                            &                    &    \\\hline
col. 9          & col. 10            & col. 11     & col. 12        & col. 13            & col. 14                    & col. 15            & col. 16 \\\hline
$n(V)$          & $\langle V\rangle$ & $\sigma(V)$ & $J$            & $N_J(40\arcs)$     & $F_{J,{\rm cor}}(40\arcs)$ & $J_{\rm cor}$      & $(V-J_{\rm cor})$\\\hline
525             & 10.139             & 0.042       & 8.905          & 3                  & 1.005                      &  8.900             & 1.239\\
276             & 11.429             & 0.117       & 9.772          & 10                 & 1.192                      &  9.581             & 1.848\\
264             & 11.800             & 0.049       &10.103          &  7                 & 1.060                      & 10.040             & 1.760\\
                &                    &             &                &                    &                            &                    &      \\\hline
col. 17         & col. 18            & col. 19     & col. 20        & col. 21            & col. 22                    & col. 23            & col. 24\\\hline
cts             &  err(cts)          & {\it HR1}   & err({\it HR1}) & $log(F_{\rm bol})$ & $R_X$                      &  err$(R_X)$        & {\it P} [d]\\\hline
0.070           & 0.013              & 0.940       & 0.120          & $-8.62$            & $-3.42$                    & 0.19               & 0.862188\\
0.082           & 0.015              & 0.230       & 0.170          & $-9.08$            & $-3.03$                    & 0.21               & 0.549447\\
0.062           & 0.013              & 0.730       & 0.170          & $-9.21$            & $-2.91$                    & 0.22               & 0.268019\\
                &                    &             &                &                    &                            &                    &\\\hline
col. 25         & col. 26            & \multicolumn{6}{c}{col. 27}\\\hline
amp             & var typ            & \multicolumn{6}{c}{ remarks}\\\hline
0.10            & rot                & \multicolumn{6}{c}{also possible EB/EC and $P=1.724375$~d}\\
0.37            & EB                 &             &                 &                   &                            &                    &\\
0.09            & rot                &             &                 &                   &                            &                    &\\ 
\hline}

The catalog of the 307 periodic variables is given in Table~1. For
illustration, we list here the first three entries of the catalog and
Fig.~5 shows a few examples of the light curves. Full Table~1 and all light
curves are available in the electronic form from the {\it Acta Astronomica
Archive (ftp://ftp.astrouw.edu.pl/acta/2013/kir\_53)}.\\ 
The consecutive columns of Table~1 contain the following data:

\begin{itemize}
\parskip=0pt \itemsep=1mm \setlength{\itemsep}{0.4mm}
\setlength{\parindent}{-1em} \setlength{\itemindent}{-1em}
\item[]Column 1 -- X-ray source designation from the RBSC catalog (1RXS J${\rm hhmmss.s}\pm{\rm ddmmss}$).
\item[]Column 2 -- ASAS designation of the optical ASAS counterpart to the  RBSC source (${\rm hhmmss}\pm{\rm ddmm.m}$). 
\item[]Column 3  -- a distance in arc seconds between the position of the ROSAT object and its ASAS counterpart (note that one pixel of the ASAS detector $\approx15\arcs$).
\item[]Columns 4--5 -- another name of the star from the SIMBAD database.
\item[]Column 6 -- spectral type from SIMBAD, if available.
\item[]Columns 7 and 8 -- $\mu_\alpha$ and $\mu_\delta$, proper motion in RA, and Dec. [mas/year].
\item[]Column 9 --  number of observations in {\it V}.
\item[]Columns 10 and 11 -- $\langle V\rangle$, mean {\it V}-magnitude with uncertainty [mag].
\item[]Column 12 -- {\it J}-magnitude of the adopted counterpart from the 2MASS catalog.
\item[]Column 13 -- number of {\it J}-band sources within 40\arcs from the ASAS object position.
\item[]Column 14 -- the relative total {\it J}-flux of all sources counted in column~13 and expressed in units of the flux from column~12.
\item[]Column 15 -- $J_{\rm cor}$, the adopted {\it J}-magnitude of the ASAS object.
\item[]Column 16 -- color index $(V-J_{\rm cor})$ (column~10 {\it minus} column~15).
\item[]Columns 17 and 18  --  number of counts per second and its error (RBSC data).
\item[]Columns 19 and 20 --  hardness ratio {\it HR1}, and its error (RBSC data).
\item[]Column 21 -- $\log F_{\rm bol}$, logarithm of the bolometric flux [ergs/s/cm$^2$].
\item[]Columns 22 and 23  -- $R_x$, logarithm of the X-ray (0.1--2.4~keV) to the bolometric flux ratio and its error.
\item[]Column 24 -- adopted variability period [days].
\item[]Column 25 -- maximum amplitude of the {\it V}-variability [mag].
\item[]Column 26 -- variability type (rot -- spotted rotator, ED -- detached binary, EB -- $\beta$~Lyr type binary, EC -- contact binary, msc -- miscellaneous).
\item[]Column 27 -- remarks
\end{itemize}

\vspace*{-11pt}
\renewcommand{\TableFont}{\footnotesize}
\MakeTableee{|c|c|c|c|l|}{12.5cm}{Previously known variable
stars with a determined or corrected period}
{\hline
\uprule
ASAS    & other name & new period & old period & remarks and\\
\dorule &            & [d]        & [d]        & references\\
\hline
\uprule
005550+5046.2 & TYC 3274-955-1    & 9.80240 & 1.65977 & EB, H09     \\
011611+2834.8 & HD 7579           & 0.998099&  --     & lt, H31     \\
024242+5027.8 & V562 Per          & 3.98440 &  --     & Hv,  P97    \\
033311+1035.9 & V1267 Tau         & 2.113   &  --     & lt, TT*     \\
051111+2813.8 & TYC 1858-529-1    & 2.0141  & 15.115  & lt, TT*, P02\\
055842-0311.0 &ASAS 055842-0311.1 & 2.202   & 32.82   & P02         \\
065847+2843.0 & 2E 1751           & 1.3175  & 1.6063  & H11         \\
110551+5151.3 & HI UMa            & 30.83   & 11.01   & KE02        \\
125533+3011.1 & NR Com            & 5.2109  & 2.5903  & N07         \\
143115+4535.7 & EE Boo            & 49.90   &  --     & Hv, P97     \\
202740+7734.3 & BD+77 779         & 56.80   &  --     & lp. S59     \\
213302+6200.2 & V430 Cep          & 7.771000&  --     & Hv, P97     \\
230604+6355.6 & GJ 9809           & 2.831000& 4.501   & KE02        \\
\hline
\noalign{\vskip3pt}
\multicolumn{5}{p{11cm}}{
Remarks: lt -- long term variability, Hv -- unsolved Hipparcos variable, 
lp -- long period variable, EB -- $\beta$~Lyr type eclipsing star, TT* -- T~Tau
type star.\newline
References: H09 -- Hoffman \etal 2009, H11 -- Hartman \etal 2011,
H31 -- Hoffmeister 1931, KE02 -- Koen and Eyer 2002, N07 -- Norton \etal
2007, P02 -- Pojmañski 2002, P97 -- Perryman \etal 1997, S59 -- Strohmeier 1959.
}}
Variability of thirteen stars listed in our catalog has already been noted
in the literature, with periods determined for seven of them. We confirm
variability of all of them. They are included into our catalog because we
determined variability periods for those six stars with unknown periods and
we revised the previously known values for the remaining seven. We also
list them separately in Table~2 with additional information.

Very few data exist on parallaxes and/or radial velocity measurements for
the investigated stars. We list them in Table~3 with some data repeated
from Table~1. It is seen from Table~3 that full kinematic data exist only
for seven stars although the value of the parallax for the star ASAS
120216+7221.7 is lower than its uncertainty.
\renewcommand{\arraystretch}{1.15}
\renewcommand{\TableFont}{\scriptsize}
\MakeTable{|c|c|r@{\hspace{0pt}}c@{\hspace{0pt}}l|r@{\hspace{0pt}}c@{\hspace{0pt}}l|r@{\hspace{0pt}}c@{\hspace{0pt}}l|r@{\hspace{0pt}}c@{\hspace{0pt}}l|r@{\hspace{0pt}}c@{\hspace{0pt}}l|}{12.5cm}{List of stars from Table~1 for which kinematical data are
available}
{\hline
\douprule
ASAS  & name & \multicolumn{3}{c|}{$\mu_\alpha$}&\multicolumn{3}{c|}{$\mu_\delta$} & \multicolumn{3}{c|}{par.} & \multicolumn{3}{c|}{$V_{\rm rad}$} & \multicolumn{3}{c|}{{\it P}} \\
\hline
\uprule
010122-0519.2 & HD 6011                 & $  14$&.&$18$ & $  -8$&.&$92$ &  7&.&83  &     &--&     &   2&.&11457 \\
021757+3115.8 & BD+30 367               & $ -44$&.&$1 $ & $ -53$&.&$5 $ &   &--&   & $ 27$&.&$25$ &   0&.&310695 \\ 
022113+4600.1 & BD+45 598               & $  44$&.&$7 $ & $ -44$&.&$3 $ &   &--&   & $ -0$&.&$8 $ &   4&.&729 \\
035720+5051.3 & HD 232862               & $  54$&.&$9 $ & $ -75$&.&$0 $ &   &--&   & $ -1$&.&$80$ &   1&.&816 \\
051111+2813.8 & 2MASS J05111053+2813504 & $   4$&.&$8 $ & $ -24$&.&$5 $ &   &--&   & $ 15$&.&$05$ &   2&.&0141 \\
053139-0327.2 & HD 294207               & $  45$&.&$2 $ & $ -52$&.&$9 $ &   &--&   & $ 27$&.&$2 $ &   6&.&642 \\  
053704+5231.4 & G 191-47                & $ 100$&.&$90$ & $-208$&.&$67$ & 27&.&61  & $ 15$&.&$  $ &   7&.&783 \\
093951-2134.3 & 2MASS J09395143-2134175 & $ -48$&.&$1 $ & $   6$&.&$2 $ &   &--&   & $ 18$&.&$3 $ &   0&.&7952 \\
100228+4434.7 & G 146-7                 & $-283$&.&$43$ & $ -90$&.&$17$ & 36&.&26  & $ 30$&.&$16$ &  14&.&92 \\
110551+5151.3 & HI UMa                  & $ -15$&.&$96$ & $  -7$&.&$69$ &  3&.&26  &     &--&     &  30&.&824 \\
120216+7221.7 & BD+73 543A              & $  11$&.&$48$ & $   7$&.&$52$ &  0&.&73  & $-54$&.&$02$ &   7&.&573 \\
125523+6953.2 & BD+70 720               & $ -16$&.&$36$ & $ -24$&.&$94$ & 13&.&77  & $-28$&.&$1 $ &   4&.&0245 \\
131300+5048.9 & LP 132-79               & $-111$&.&$11$ & $  95$&.&$83$ & 11&.&12  &      &--&    &   0&.&38995 \\
134536+6548.2 & HD 120163               & $ -28$&.&$55$ & $  17$&.&$51$ &  3&.&18  &      &--&    &  14&.&03 \\
143115+4535.7 & EE Boo                  & $ -53$&.&$28$ & $  15$&.&$24$ &  3&.&60  &      &--&    &  49&.&9 \\
150804+6943.9 & NLTT 39471              & $-147$&.&$80$ & $  77$&.&$08$ &  7&.&36  &      &--&    &   8&.&628 \\
204128+5725.8 & LTT 16050               & $  97$&.&$99$ & $ 206$&.&$17$ & 41&.&26  &      &--&    &  14&.&118 \\
213240+3604.7 & HD 205173               & $  16$&.&$6 $ & $  -1$&.&$8 $ &   &--&   & $ -6$&.&$2 $ & 101&.&8 \\
213302+6200.2 & V430 Cep                & $ 371$&.&$28$ & $ 194$&.&$39$ & 43&.&03  & $-11$&.&$14$ &   7&.&771 \\
221601-1411.0 & 2MASS J22160063-1411022 & $  50$&.&$22$ & $  38$&.&$71$ & 17&.&71  & $-24$&.&$4 $ &   0&.&64573\\
230236+7630.3 & HD 218028               & $ 151$&.&$74$ & $  45$&.&$91$ & 14&.&97  &     &--&     &   2&.&598\\
230605+6355.6 & GJ 9809                 & $ 171$&.&$46$ & $ -58$&.&$55$ & 40&.&81  & $-23$&.&$5 $ &   2&.&831\\
234008-0228.9 & BD-03 5686              & $ -34$&.&$0 $ & $ -50$&.&$8 $ &   &--&   & $ 66$&.&$3 $ &   2&.&8968\\
235451+3831.6 & 2MASS J3545147+3831363  & $-130$&.&$0 $ & $ -86$&.&$0 $ &   &--&   & $  5$&.&$  $ &   4&.&757 \\
\hline
\noalign{\vskip4pt}
\multicolumn{17}{p{12.5cm}}{Column descriptions: 
$\mu_\alpha $, $\mu_\delta$ -- proper motion in right ascension and
declination [mas/year], par. -- heliocentric parallax [mas], $V_{\rm rad}$ --
radial velocity [km/s], $P$ -- variability period [d]}}

\vspace*{-9pt}
\Section{Summary}
\vspace*{-5pt}
We present the results of the second part of periodic variability search
among stars observed within the photometric survey ASAS and showing a high
level of the coronal activity as measured by X-ray flux. This time the
Northern part of the survey was analyzed. A catalog with the data on 307
new periodic variables (including a few previously known variables for
which the variability period was unknown or in error) is given.

We classified 266 stars as spotted rotators. Among eclipsing stars we found
10 contact binaries, 12 eclipsing binaries with deformed components but not
in contact, and 18 detached systems.

Most of the stars have $V-J$ colors and spectral classification consistent
with a presence of the outer convection zone and coronal X-ray
emission. However, we also identified a few objects with different
properties. In particular, the hottest object in the sample is likely a
cataclysmic variable, whereas the source of X-ray emission from the reddest
star in our list, ASAS 205846+1417.8, is somewhat puzzling. The star may be
an M-type giant due to a small proper motion and a large bolometric
luminosity, but the stars of this type are not known to be active coronal
sources.

\Acknow{We are particularly grateful to Prof.\ Grzegorz Pojmañski for help
in using ASAS data. This research was partly supported by the National
Science Centre under the grant DEC-2011/03/B/ST9/03299. We acknowledge
the use of the SIMBAD database, operated at CDS, Strasbourg, France. This
publication makes use of data products from the Two Micron All Sky
Survey, which is a joint project of the University of Massachusetts and
the Infrared Processing and Analysis Center/California Institute of
Technology, funded by the National Aeronautics and Space Administration
and the National Science Foundation.}

\end{document}